\documentclass[preprint,aps]{revtex4} 
\usepackage{graphicx}

\begin{document}
\title{Relational physics with real rods and clocks 
and the measurement problem of quantum mechanics}

\author{Rodolfo Gambini$^{1}$ 
and Jorge Pullin$^{2}$} 
\affiliation {1. Instituto de F\'{\i}sica,
Facultad de Ciencias, Igu\'a 4225, esq. Mataojo, Montevideo, Uruguay. \\
2. Department of Physics and Astronomy, Louisiana State
University, Baton Rouge, LA 70803-4001} 

\date{July 6th 2006; revised Dec. 5th 2006}

\begin{abstract}
The use of real clocks and measuring rods in quantum mechanics implies
a natural loss of unitarity in the description of the theory. We briefly review
this point and then discuss the implications it has for the measurement
problem in quantum mechanics. The intrinsic loss of coherence allows to
circumvent some of the usual objections to the measurement process as
due to environmental decoherence. 
\end{abstract}
\maketitle

\section{Introduction}

The Copenhagen interpretation of quantum mechanics requires the
existence of a classical macroscopic realm in order to explain the
measurement process. In spite of this long held view, it is becoming
increasingly clear that quantum mechanics should be understood
entirely as a standalone quantum paradigm without having to refer to
an external classical world. The issue is becoming more pressing as
there exists a growing number of experiments showing the existence of
superpositions of macroscopically distinct quantum states.

Even though we tend to consider that quantum mechanics is a universal
scheme able to describe all the physical phenomena, it remains to be
shown that quantum mechanics is sufficient to explain our observation
of a classical world. There is an extended consensus among physicists
that environment-induced decoherence may play an important role in the
solution of the measurement problem of quantum mechanics.  It allows
us to understand how the interaction with the environment induces a
local suppression of interference between a set of preferred states,
associated with the ``pointer basis''. More precisely, a) decoherence
induces a fast suppression of the interference terms of the reduced
matrix describing the system coupled to the measurement device, and b)
it selects a preferred set of states that are robust in spite of their
interaction with the environment. These facts are a direct consequence
of the standard unitary time evolution of the total system-environment
composition and therefore the global phase coherence is not destroyed
but simply transferred from the system to the environment.

We have recently noticed that quantum mechanics also leads to other
kinds of loss of coherence due to the quantum effects in real clocks
\cite{njp,prl}. In fact, as ordinarily formulated, quantum mechanics
involves an idealization. That is, the use of a perfect classical
clock to measure times. Such a device clearly does not exist in
nature, since all measuring devices are subject to some level of
quantum fluctuations. The equations of quantum mechanics, when cast in
terms of the variable that is really measured by a clock in the
laboratory, differ from the traditional Schroedinger description.
Although this is an idea that arises naturally in ordinary quantum
mechanics, it is of paramount importance when one is discussing
quantum gravity. This is due to the fact that general relativity is a
generally covariant theory where one needs to describe the evolution
in a relational way. This problem is most obvious in the context of
quantum cosmology. Although space-time is usually described in terms
of fiduciary non-observable quantities like the value of the
components of the metric in a coordinate system, the final physical
questions end up being what are the values of certain physical
quantities when others physical quantities taken as clocks and
measuring rods take certain values at the same fiducial coordinate
point. Contrary to what happens with the environment-induced
decoherence this new type of decoherence is associated to a
non-unitary evolution in the physical time.  The origin of the lack of
unitarity is the fact that in quantum mechanics the determination of
the state of a system is only possible by repeating an experiment.  If
one uses a real clock, which has thermal and quantum fluctuations,
each experimental run will correspond to a different value of the
evolution parameter. The statistical prediction will therefore
correspond to an average over several intervals, and therefore its
evolution cannot be unitary.  As has been observed by Salecker and
Wigner \cite{wigner} and more recently by Ng \cite{ng}, quantum
mechanics and general relativity impose fundamental limitations to how
good a clock can be and therefore any physical system will suffer loss
of quantum coherence.  This is a fundamental inescapable limit. One
may argue about the level at which the limitation arises, but the
existence of a limitation at some level is non-controversial. A pure
state inevitably will become a mixed state due to the impossibility of
having a perfect classical clock in nature. In that sense it is
important to stress that the strict application of quantum mechanics
when the quantum nature of the clocks is taken into account leads to a
modification of the Schroedinger evolution (a similar consideration of
real ``measuring rods'' for space adds further corrections to quantum
field theory as well).

Due to the extreme accuracy that real clocks can reach this effect is
very small. A rough measure of the effect is given by the off diagonal
terms in the evolution of the density matrix for a physical system in
the energy eigenbasis. Taking into account the previously mentioned
limits for time measurements, one gets that the off diagonal terms are
depressed by the exponential of $\omega^2 t_{\rm Planck}^{4/3}
t^{2/3}$ , where $\omega$ is the Bohr frequency associated to the levels,
$t_{\rm Planck}$ is Planck's time and and $t$ the elapsed time. So we
conclude that any physical system that we study in the lab will suffer
loss of quantum coherence at least at this rate. This is a fundamental
inescapable limit. A pure state inevitably will become a mixed state
due to the impossibility of having a perfect classical clock in
nature. One can show that this effect, in absence of
environment-decoherence, would not be relevant for most ``Schroedinger
cat'' experiences involving less than $10^{10}$ atoms.

The aim of the present paper is to present a first discussion of the
interplay between these two types of decoherence and their
consequences for the problem of measurements in quantum mechanics. In
fact, standard decoherence by the environment leads to a reduced
density matrix for the system coupled with the measurement apparatus
that is approximately diagonal, this density matrix describes all the
information that can be extracted by an observer when the correlations
with the environment are ignored. That implies that a measurement of
an observable that only pertains to the system plus the measurement
device cannot discriminate between the total pure state and a mixed
state. However, as it was extensively discussed by d'Espagnat
\cite{despa} the formal identity between the reduced density matrix
and a mixed-state density matrix is frequently misinterpreted as
implying that the system is in a mixed state. As the system is
entangled with the environment the total system is still described by
a pure state and no individual definite state or set of possible
states may be attributed to a portion of the total system. As it is
well known, joint measurements of the environment and the system will
always allow us, in principle, to distinguish between the reduced and
mixed state density matrices.

The combined effect of these two forms of decoherence could allow to
understand the physical transition from a reduced density matrix to a
mixed state. In fact, the precise unitary evolution of the total
system is broken by the clock-induced decoherence destroying the
correlations. What remains to be studied is whether this effect is
sufficiently fast to avoid any possibility of distinguishing, not only
for all practical purposes but also on theoretical basis, between
these two kinds of density matrices.

Understanding this transition may be crucial for determining which
interpretations are compatible with quantum mechanics. In fact if 
there is not any deviation from the unitary dynamics, it would be
compelling to consider the universal validity of the evolution of any
system with the Schroedinger equation and therefore to single out a
relative state-type Everett interpretation. In that case the resulting
appearance of a classical regime should be considered as an apparent
effect for observers in each branch.

Summarizing, when one takes into account that we are using real clocks,
the standard postulates of quantum mechanics lead to the unexpected
fact that the evolution in terms of a real time clock is not exactly
unitary. This loss of exact unitarity may help to understand the
transition of the reduced density matrix to a mixed state density
matrix and consequently to put on equal footing the relative state
interpretation with realistic interpretations of the measurement
problem in which the observation of the system plus the apparatus in a
definite state is possible.

In the remaining sections of the paper we will give more details
concerning these ideas. In the next section we outline how the loss of
coherence arises in ordinary quantum mechanics when one considers the
use of real clocks. In section 3 we discuss the application to the
measurement problem. In section 4 we discuss the open problems that
must be tackled before the proposed mechanism can be considered a
completely satisfactory solution to the problem of measurement in
quantum mechanics.

\section{Quantum mechanics with real clocks}

We proceed to describe how is the appearance of ordinary quantum
mechanics when cast in terms of real clocks.  Given a physical
situation of interest described by a (multi-dimensional) phase space
$q,p$, we start by choosing a ``clock''. By this we mean a physical
quantity (more precisely a set of quantities, like when one chooses a
clock and a calendar to monitor periods of more than a day) that we
will use to keep track of the passage of {\it time}. An example of
such a variable could be the angular position of the hand of an analog
watch. Let us denote it by $T(q,p)$. We then identify some physical
variables that we wish to study as a function of time.  We shall call
them generically $O(q,p)$ (``observables''). We then proceed to
quantize the system by promoting all the observables and the clock
variable to self-adjoint quantum operators acting on a Hilbert
space. The latter is defined once a well defined inner product is
chosen in the set of all physically allowed states.  Usually it
consists of squared integrable functions $\psi(q)$.

Notice that we are not in any way modifying quantum mechanics. We
assume that the system has an evolution in terms of an external
parameter $t$, which is a classical variable, given by a Hamiltonian and
with operators evolving with Heisenberg's equations (it is easier to
present things in the Heisenberg picture, though it is not mandatory
to use it for our construction). Then the standard rules of quantum 
mechanics and its probabilistic nature apply.

We will call the eigenvalues of the
``clock'' operator $T$ and the eigenvalues of the ``observables'' $O$.
We define the projector associated to the measurement of the time variable
within
the interval $[T_0-\Delta T,T_0+\Delta T]$,
\begin{equation}
P_{T_0}(t) =\int_{T_0-\Delta T}^{T_0+\Delta T} dT \sum_k |T,k,t><T,k,t|
\end{equation}
where $k$ denotes the eigenvalues of the operators that form a
complete set with $\hat{T}$ (the eigenvalues can have continuous or
discrete spectrum, in the former case the sum should be replaced by an
integral).  We have assumed a continuous spectrum for $T$ therefore
the need for the integral over an interval on the right hand side.
The interval $\Delta T$ is assumed to be very small compared to any of
the times intervals of interest in the problem, in particular the time
separating two successive measurements. Similarly we introduce a
projector associated with the measurement of the observable $O$,
\begin{equation}
P_{O_0}(t) =\int_{O_0-\Delta O}^{O_0+\Delta O} dO \sum_j |O,j,t><O,j,t|
\end{equation}
with $j$ the eigenvalues of a set of operators that form a complete
set with $\hat{O}$. These projectors have the usual properties, i.e.,
$P_a(t)^2 = P_a(t)$, $\sum_a P_a(t)=1, \forall t$ and
$P_a(t)P_{a'}(t)= 0$ if the intervals surrounding $a$ and $a'$ do not
have overlap.

We would like now to ask the question ``what is the probability that
the observable $O$ take a given value $O_0$ given that the clock
indicates a certain time $T_0$ ''. One here is assuming therefore 
that one has access to a physical clock and that underlying behind
the framework is a foliation of space-time constructed using an
inaccessible idealized clock $t$ so the clock reading and the 
apparatus measurement occur at the same idealized time $t$. 
Such question is embodied in the
conditional probability,
\begin{equation}\label{condprob}
{\cal P}\left( O\in [O_0-\Delta O,O_0+\Delta O]|T \in 
[T_0-\Delta T,T_0+\Delta T]\right) =\lim_{\tau\to\infty}
{\int_{-\tau}^{\tau} dt\, {\rm Tr}\left(P_{O_0}(t) P_{T_0}(t)\rho 
P_{T_0}(t)\right) 
\over 
{\int_{-\tau}^{\tau} dt\,{\rm Tr}\left(P_{T_0}(t)\rho \right)}}
\end{equation}
where we have used the properties of both projectors and the integrals
over $t$ in the right hand side are taken over all its possible
values.  The reason for the integrals is that we do not know for what
value of the external ideal time $t$ the clock will take the value
$T_0$. In this expression $\rho$ is the density matrix of the system.
One has to take some obvious cares, like for instance to choose a
clock variable (or set of variables) that do not take twice the same
value during the relevant lifetime of the experiment one is
considering. It should be noted that in this context the role of the
fiducial time $t$ is similar to that of the use of coordinates to
describe space-time in general relativity. At the end of the day
physical questions are embodied in relational measurements of
quantities and no specific direct knowledge of the fiduciary
parameters is needed. One may philosophically speculate as to why to
assume that the evolution is unitary in the fiduciary parameter
$t$. Here one could note that ordinary experience with quantum
mechanics with real clocks suggests that evolution is at worse very
approximately unitary. It is therefore natural to assume it is exactly
unitary in the parameter $t$ and approximately unitary in physical not
absolutely accurate clocks $T$ one may consider.

The above expression is general, it will apply to any choice of
``clock'' and ``system'' variables we make. The relational evolution
of the conditional probabilities will be complicated and will bear
little resemblance to the usual evolution of probabilities in ordinary
quantum mechanics unless we make a ``wise'' selection of the clock and
system variables. What we mean by this is that we would like to choose
as clock variables a subsystem that interacts little with the system
we want to study and that behaves semi-classically with small quantum
fluctuations. Namely, the physical clock will be correlated with the
ideal time in such a way to produce the usual notion of time. In
such a regime one expects to recover ordinary Schroedinger evolution
(plus small corrections) even if one is using a ``real'' clock. Let us
consider such a limit in detail. We will assume that we divide the
density matrix of the whole system into a product form between clock
and system, $\rho =\rho_{\rm cl}\otimes \rho_{\rm sys}$ and the
evolution will be given by a unitary operator also of product type
$U=U_{\rm cl}\otimes U_{\rm sys}$.

Up to now we have considered the quantum states as described by a 
density matrix at a time $t$. Since the latter is unobservable, we 
would like to shift to a description where we have density matrices
as functions of the observable time $T$. To do this, we recall the 
expression for the usual probability in the Schroedinger representation
of measuring the value $O$ at a time $t$,
\begin{equation}
  {\cal P}\left(O|t\right)\equiv {{\rm Tr}\left(P_{O}(0)\rho(t)\right)
\over {\rm Tr}\left(\rho(t)\right)}\label{ordinary}
\end{equation}
where the projector is evaluated at $t=0$ since in the Schroedinger
picture operators do not evolve. We would like to get a similar expression
in terms of the real clock. To do this we consider the conditional
probability (\ref{condprob}), and make explicit the separation between
clock and system,
\begin{eqnarray} \label{condprob2}
&&{\cal P}\left( O\in [O_0\pm\Delta O]|T \in 
[T_0\pm\Delta T]\right) =\\
&&\lim_{\tau\to\infty}
{\int_{-\tau}^{\tau} dt\,{\rm Tr}\left(U_{\rm sys}(t)^\dagger P_O(0)
U_{\rm sys}(t) U_{\rm cl}(t)^\dagger P_T(0) 
U_{\rm cl}(t)\rho_{\rm sys}\otimes 
\rho_{\rm cl}\right) 
\over 
{\int_{-\tau}^{\tau} dt\,{\rm Tr}
\left(P_T(t)\rho_{\rm cl} \right) 
{\rm Tr}\left(\rho_{\rm sys}\right)}}\nonumber\\
&=&
\lim_{\tau\to\infty}
{\int_{-\tau}^{\tau} dt\,{\rm Tr}\left(U_{\rm sys}(t)^\dagger P_O(0)
U_{\rm sys}(t) \rho_{\rm sys}\right){\rm Tr}
\left(U_{\rm cl}(t)^\dagger P_T(0) U_{\rm cl}(t) 
\rho_{\rm cl}\right) 
\over 
{\int_{-\tau}^{\tau} dt\,
{\rm Tr}\left(P_T(t)\rho_{\rm cl} \right) {\rm Tr}
\left(\rho_{\rm sys}\right)}}.
\end{eqnarray}

We define the probability density 
that the resulting measurement of the clock
variable takes the value $T$ when the ideal time
takes the value  $t$,
\begin{equation}
{\cal P}_t(T) \equiv {{\rm Tr}\left(P_T(0) U_{\rm cl}(t)
\rho_{\rm cl} U_{\rm cl}(t)^\dagger\right)\over
\int_{-\infty}^\infty dt\,{\rm Tr}\left(P_T(t) \rho_{\rm cl}\right)},
\end{equation}
and notice that $\int_{-\infty}^\infty dt {\cal P}_t(T)=1$. We now
define the evolution of the density matrix,
\begin{equation}
\rho(T) \equiv \int_{-\infty}^\infty dt U_{\rm sys}(t) \rho_{\rm sys} 
 U_{\rm sys}(t)^\dagger {\cal P}_t(T)
\end{equation}
where we dropped the ``sys'' subscript in the left hand side since
it is obvious we are ultimately interested in the density matrix of
the system under study, not that of the clock. Noting that
\begin{equation}
  {\rm Tr}\left(\rho(T)\right)=\int_{-\infty}^\infty dt\, {\cal P}_t(T) 
{\rm Tr}\left(\rho_{\rm sys}\right)={\rm Tr}\left(\rho_{\rm sys}\right),
\end{equation}
one can equate the conditional probability (\ref{condprob2}) to the
ordinary probability of quantum mechanics (\ref{ordinary}),
\begin{equation}
{\cal P}\left( O\in [O_0\pm\Delta O]|T \in 
[T_0\pm\Delta T]\right) =
{{\rm Tr}\left(P_{T_O}(0)\rho(T)\right)
\over {\rm Tr}\left(\rho(T)\right)},
\end{equation}
(and we omit the integrals needed to make the expression precise 
in the case of continuous spectra for brevity). 
We have
therefore ended with the standard probability expression with an
``effective'' density matrix in the Schroedinger picture given by 
$\rho(T)$. By its very definition, it is immediate to see that in
the resulting evolution unitarity is lost, since one ends up with
a density matrix that is a superposition of density matrices 
associated with different $t$'s and that each evolve unitarily 
according to ordinary quantum mechanics.

Now that we have identified what will play the role of a density
matrix in terms of a ``real clock'' evolution, we would like to see
what happens if we assume the ``real clock'' is behaving
semi-classically. To do this we assume that ${\cal P}_t(T) =f(T,T_{\rm
  max}(t))$, where $f$ is a function that decays very rapidly for
values of $T$ far from the maximum of the probability distribution
$T_{\rm max}$. To make the expressions as simple as possible, let
us assume that $T_{\rm max}(t)=t$, i.e. the peak of the probability
distribution is simply at $t$. More general dependences can of course
be considered, altering the formulas minimally (for a more complete 
treatment see \cite{njp}). We will also assume
that we can approximate $f$ reasonably well by a Dirac delta, 
namely,
\begin{equation}
f(T,t)=\delta(T-t)+ a(T)\delta'(T-t)+b(T)\delta''(T-t)+\ldots,
\end{equation}
where the first term has a unit coefficient so the integral of the probability
is unit and we assume $b(T)>0$ so it represents extra width with respect to
the Dirac delta.

We now consider the evolution of the density matrix,
\begin{equation}
\rho(T)=\int_{-\infty}^\infty dt\,\rho_{\rm sys}(t) {\cal P}_t(T)=
\int_{-\infty}^\infty dt\,\rho_{\rm sys}(t) f(T,t)
\end{equation}
and associating a Hamiltonian with the evolution operator $U(t)=\exp(iHt)$,
we get,
\begin{equation}
\rho(T)=\rho_{\rm sys}(T)+a(T) [H,\rho_{\rm sys}(T)]-b(T) 
[H,[H,\rho_{\rm sys}(t)]],
\end{equation}
and we notice that there would be terms involving further commutators if
we had kept further terms in the expansion of $f(T,t)$ in terms of the
Dirac deltas.

We can now consider the time derivative of this expression, and get,
\begin{equation}
{\partial \rho(T)\over \partial T} =i\left(-1
+{\partial a(T)\over \partial T}\right) 
[H,\rho(T)] +\left(a(T)-{\partial b(T)\over 
\partial T}\right) [H,[H,\rho(T)]].
\end{equation}
If we had considered a symmetric distribution (it is natural to
consider such distributions since on average one does not expect an
effect that would lead systematically to values grater or smaller than
the mean value), we see that one would have obtained the traditional
evolution to leading order plus a corrective term,
\begin{equation}
{\partial \rho(T)\over \partial T} =i [\rho(T),H] +\sigma(T) [H,[H,\rho(T)]].
\end{equation}
and the extra term is dominated by the rate of change of the width of 
the distribution $\sigma(T)=\partial b(T)/\partial T$. 

An equation of this form has been considered in the context of
decoherence due to environmental effects, it is called the Lindblad
equation \cite{lindblad},
\begin{equation}
\frac{d}{dt}\rho=-i[H,\rho]-{\cal D}(\rho)\label{rho},
\end{equation}
with
\begin{equation} {\cal D}(\rho)=\sum_n[D_n,[D_n,\rho]], \;\;\;
D_n=D_n^{\dagger},\;\;\; [D_n,H]=0,
\end{equation}
and in our case there is only one $D_n$ that is non-vanishing and it
coincides with $H$. This is a desirable thing, since it implies that
conserved quantities are automatically preserved by the modified
evolution.  Other mechanisms of decoherence coming from a different
set of effects of quantum gravity have been criticized in the past
because they fail to conserve energy
\cite{hag}. It should be noted that Milburn arrived at a similar
equation as ours from different assumptions \cite{milburn}. Egusquiza,
Garay and Raya derived a similar expression from considering
imperfections in the clock due to thermal fluctuations
\cite{egusquiza}. It is to be noted that such effects will occur in
addition to the ones we discuss here.

What is the effect of the extra term? To study this, let us pretend for
a moment that $\sigma(T)$ is constant. That is, the distribution in the
clock variable has a width that grows linearly with time. In that case,
the evolution equation is exactly solvable. If we consider a system
with energy levels, the elements of the density matrix in the energy
eigenbasis is given by,
\begin{equation}
  \rho(T)_{nm} = \rho_{nm}(0) e^{-i\omega_{nm} T} e^{-\sigma \omega_{nm}^2 T}
\end{equation}
where $\omega_{nm}=\omega_n-\omega_m$ is the Bohr frequency
corresponding to the levels $n,m$. We therefore see that the
off-diagonal elements of the density matrix go to zero exponentially
at a rate governed by $\sigma$, i.e. by how badly the clock's
wavefunction spreads. It is clear that a pure state is eventually
transformed into a completely mixed state (``proper mixture'' in
d'Espagnat's terminology) by this process.

The origin of the lack of unitarity is the fact that definite
statistical predictions are only possible by repeating an experiment.
If one uses a real clock, which has thermal and quantum fluctuations,
each experimental run will correspond to a different value of the
evolution parameter. The statistical prediction will therefore
correspond to an average over several intervals, and therefore its
evolution cannot be unitary. 

In a real experiment, there will be decoherence in the system under
study due to interactions with the environment, that will be superposed
on the effect we discuss. Such interactions might be reduced by
cleverly setting up the experiment. The decoherence we are discussing
here however, is completely determined by the quality of the clock used.
It is clear that if one does experiments in quantum mechanics with 
poor clocks, pure states will evolve into mixed states very rapidly.
The effect we are discussing can therefore be magnified arbitrarily
simply by choosing a lousy clock. This effect has actually been observed
experimentally in the Rabi oscillations describing the exchange of
excitations between atoms and field \cite{Br}.

We have established that when we study quantum mechanics with a
physical clock (a clock that includes quantum fluctuations), unitarity
is lost, conserved quantities are still preserved, and pure states
evolve into mixed states. The effects are more pronounced the worse
the clock is.  Which raises the question: is there a fundamental
limitation to how good a clock can be? This question was first
addressed by Salecker and Wigner \cite{wigner}. Their reasoning went
as follows: suppose we want to build the best clock we can. We start
by insulating it from any interactions with the environment.  An
elementary clock can be built by considering a photon bouncing between
two mirrors. The clock ``ticks'' every time the photon strikes one of
the mirrors. Such a clock, even completely isolated from any
environmental effects, develops errors. The reason for them is that by
the time the photon travels between the mirrors, the wavefunctions of
the mirrors spread. Therefore the time of arrival of the photon
develops an uncertainty.  Salecker and Wigner calculated the
uncertainty to be $\delta t \sim \sqrt{t/M}$ where $M$ is the mass of
the mirrors and $t$ is the time to be measured (we are using units
where $\hbar=c=1$ and therefore mass is measured in 1/second). The
longer the time measured the larger the error. The larger the
mass of the clock, the smaller the error.

So this tells us that one can build an arbitrarily accurate clock just
by increasing its mass. However, Ng and Van Damme \cite{ng} pointed
out that there is a limit to this. Basically, if one piles up enough
mass in a concentrated region of space one ends up with a black hole.
Some readers may ponder why do we need to consider a concentrated
region of space. The reason is that if we allow the clock to be more
massive by making it bigger, it also deteriorates its performance.
For instance, in the case of two mirrors and a photon, if one makes
the mirror big, there will be uncertainty in its position due to
elastic effects like sound waves traveling across it, which will
negate the effect of the additional mass (see the discussion in
\cite{ngotro} in response to \cite{baez}).
  
A black hole can be thought of as a clock (as we will see it turns
out to be the most accurate clock one can have).  It has normal modes of
vibration that have frequencies that are of the order of the light
travel time across the Schwarzschild radius of the black hole.  (It is
amusing to note that for a solar sized black hole the frequency is in
the kilohertz range, roughly similar to that of an ordinary bell). The
more mass in the black hole, the lower the frequency, and therefore
the worse its performance as a clock. This therefore creates a tension
with the argument of Salecker and Wigner, which required more mass to
increase the accuracy. This indicates that there actually is a ``sweet
spot'' in terms of the mass that minimizes the error. Given a time to
be measured, light traveling at that speed determines a distance, and
therefore a maximum mass one could fit into a volume determined by
that distance before one forms a black hole.  That is the optimal
mass.  Taking this into account one finds that the best accuracy one
can get in a clock is given by $\delta T \sim T_{\rm Planck}^{2/3}
T^{1/3}$ where $T_{\rm Planck}=10^{-44}s$ is Planck's time and $T$ is
the time interval to be measured. This is an interesting result. On
the one hand it is small enough for ordinary times that it will not
interfere with most known physics. On the other hand is barely big
enough that one might contemplate experimentally testing it, perhaps
in future years.

With this absolute limit on the accuracy of a clock we can quickly
work out an expression for the $\sigma(T)$ that we discussed in the
previous section \cite{prl,piombino}. It turns out to be $\sigma(T)=
\left({T_{\rm Planck}}\over{T_{\rm max}-T}\right)^{1/3}T_{\rm
  Planck}$. With this estimate of the absolute best accuracy of a
clock, we can work out again the evolution of the density matrix for a
physical system in the energy eigenbasis. One gets
\begin{equation}
  \rho(T)_{nm} = \rho_{nm}(0) 
e^{-i\omega_{nm} T} e^{-\omega_{nm}^2  T_{\rm Planck}^{4/3} T^{2/3}}.
\end{equation}

So we conclude that {\em any} physical system that we study in the lab
will suffer loss of quantum coherence at least at the rate given by
the formula above. This is a fundamental inescapable limit. A pure
state inevitably will become a mixed state due to the impossibility of
having a perfect classical clock in nature.

Given these conclusions, one can ask what are the prospects for
detecting the fundamental decoherence we propose. If one would like to
observe the effect in the lab one would require that the decoherence
manifest itself in times of the order of magnitude of hours, perhaps
days at best. That requires energy differences of the order of
$10^{10}eV$ in the Bohr frequencies of the system. Such energy
differences can only be achieved in ``Schroedinger cat'' type
experiments.  Among the best candidates today are Bose--Einstein
condensates, which can have $10^6$ atoms in coherent states. These
states can have energy differences for which the fundamental
decoherence exponents become of order unity only in times larger than
the age of the universe. These effects could be observed sooner if one
could build larger coherent states, or if loss of coherence could be
monitored with high levels of precision. But the challenge of
eliminating faster acting environmental decoherence effects is huge
and at present it remains unclear if any experiment could be proposed
in the near future that could detect the fundamental decoherence.

A point that could be raised is that atomic clocks currently have an
accuracy that is less than a decade of orders of magnitude worse than
the absolute limit we derived in the previous section. Couldn't
improvements in atomic clock technology actually get better than our
supposed absolute limit? This seems unlikely. When one studies in
detail the most recent proposals to improve atomic clocks, they
require the use of entangled states \cite{atomic} that have to remain
coherent. Our effect would actually prevent the improvement of atomic
clocks beyond the absolute limit!

Another point to be emphasized is that our approach has been quite
naive in the sense that we have kept the discussion entirely in terms
of non-relativistic quantum mechanics with a unique time across space.
It is clear that in addition to the decoherence effect we discuss
here, there will also be decoherence spatially due to the fact that
one cannot have clocks perfectly synchronized across space and also
that there will be fundamental uncertainties in the determination of
spatial positions (``use of real measuring rods'').  
We have not studied this in great detail yet, but it
appears that this type of decoherence could be even more promising 
from the point of view of experimental detection (see \cite{SiJa}).
For a brief discussion of the possible effects see \cite{spatial}.

It is interesting to notice that the presence of fundamental loss
of coherence has implications for the black hole information puzzle.
We will not expand on this issue here, but we refer the reader to 
our treatment of this issue in references \cite{piombino,prl}.
Implications for quantum computing were also discussed in \cite{qc}.

\section{Implications for the measurement problem of quantum mechanics}

A potential conceptual application of the fundamental decoherence that
we discussed that has not been exploited up to now is in connection
with the measurement problem in quantum mechanics.  The latter is
related to the fact that in ordinary quantum mechanics the measurement
apparatus is assumed to be always in an eigenstate after a measurement
has been performed.  The usual explanation \cite{Schlossauser} for
this is that there exists interaction with the environment. This
selects a preferred basis, i.e., a particular set of quasi-classical
states often referred to as ``pointer states'' that are robust, in the
sense of retaining correlations over time inspite of their immersion
in the environment. These states are determined by the form of the
interaction between the system and its environment and correspond to
the classical states of our everyday experience.  Decoherence then
quickly damps superpositions between the localized preferred states
when only the system is considered. This is taken as an explanation of
the appearance to a local observer of a ``classical'' world of
determinate, ``objective'' (robust) properties.

The main problem with such a point of view is how is one to
interpret the local suppression of interference in spite of the
fact that the total state describing the system-environment
combination retains full coherence. One may raise the question 
whether retention of the full coherence could ever lead to 
empirical conflicts with the ascription of definite values to 
macroscopic systems. The usual point of view is that it would
be very difficult to reconstruct the off diagonal elements of
the density matrix in practical circumstances. However, at least
as a matter of principle, one could indeed reconstruct such
terms (the evolution of the whole system remains unitary
\cite{omnes}).

Our mechanism of fundamental decoherence could contribute to
the understanding of this issue. In the usual system-environment
interaction the off-diagonal terms of the density matrix oscillate
as a function of time. Since the environment is usually considered
to contain a very large number of degrees of freedom, the common
period of oscillation for the off-diagonal terms to recover 
non-vanishing values is very large, in many cases larger than
the life of the universe. This allows to consider the problem
solved in practical terms. When one adds in the effect we 
discussed, since it suppresses exponentially the off-diagonal
terms, one never has the possibility that the off-diagonal
terms will see their initial values restored, no matter how
long one waits. 

To analyze the implications of the use of real clocks in the
measurement problem, we will analyze an example. In spite of the
universality of the loss of coherence we introduced, it must be
studied in specific examples of increasing level of realism. The
simplest example we can think of is due to Zurek \cite{zurek}.  This
simplified model does not have all the effects of a realistic one, yet
it exhibits how the information is transferred from the measuring
apparatus to the environment. The model consists of taking a spin
one-half system that encodes the information about the microscopic
system plus the measuring device. A basis in its two dimensional
Hilbert space will be denoted by $\{|+>,|->\}$. The environment is
modeled by a bath of many similar two-state systems called
atoms. There are $N$ of them, each denoted by an index $k$ and with
associated two dimensional Hilbert space $\{|+>_k,|->_k\}$. The
dynamics is very simple, when there is no coupling with the
environment the two spin states have the same energy, which is taken
to be $0$.  All the atoms have zero energy as well in the absence of
coupling. The whole dynamics is contained in the coupling, given by
the following interaction Hamiltonian
\begin{equation}
H_{\rm int} = \hbar \sum_k \left( g_k \sigma_z \otimes \sigma_z^k \otimes
\prod_{j\neq k} I_j\right).
\end{equation}
In this notation $\sigma_z$ is analogous to a Pauli spin matrix.  It
has eigenvalues $+1$ for the spin eigenvector $|+>$ and $-1$ for
$|->$; it acts as the identity operator on all the atoms of the
environment. The operators $\sigma^k_z$ are similar, each acts like a
Pauli matrix on the states of the specific atom $k$ and as the
identity upon all the other atoms and the spin. $I_j$ denotes the
identity matrix acting on atom $j$ and $g_k$ is the coupling constant
that has dimensions of energy and characterizes the coupling energy of
one of the spins $k$ with the system. In spite of the abstract
character of the model, it can be thought of as providing a sketchy
model of a photon propagating in a polarization analyzer.

Starting from a normalized initial state
\begin{equation}
|\Psi(0)> = \left(a|+> + b|->\right) \prod_{k=1}^N \otimes \left[
\alpha_k|+>_k +\beta_k |->_k \right],
\end{equation}
it is easy to solve the Schroedinger equation and one gets for the
state at the time $t$,
\begin{eqnarray}
|\Psi(t)> 
&=& a |+> \prod_{k=1}^N\otimes\left[
\alpha_k\exp\left(ig_k t\right)|+>_k 
+ \beta_k \exp\left(-ig_k t\right)|->\right]\\
&&+ b |-> 
\prod_{k=1}^N\otimes\left[
\alpha_k\exp\left(-ig_k t\right)|+>_k 
+ \beta_k \exp\left(ig_k t\right)|->\right].\nonumber
\end{eqnarray}

Writing the complete density operator $\rho(t) = |\Psi(t)><\Psi(t)|$,
one can take its trace over the environment degrees of freedom to get
the reduced density operator,
\begin{equation}
\rho_c(t) = |a|^2 |+><+| + |b|^2 |-><-| + z(t) ab^* |+><-| 
+z^*(t) a^* b|-><+|,
\end{equation}
where
\begin{equation}
z(t) = \prod_{k=1}^N \left[\cos\left(2g_k t\right)+i\left(|\alpha_k|^2
-|\beta_k|^2\right) \sin\left(2 g_k t\right)\right].
\end{equation}

The complex number $z(t)$ controls the value of the non-diagonal
elements. If this quantity vanishes the reduced density matrix
$\rho_c$ would correspond to a totally mixed state (``proper
mixture''). That would be the desired result, one would have several
classical outcomes with their assigned probabilities. However,
although the expression we obtained vanishes quickly assuming the
$\alpha$'s and $\beta$'s take random values, it behaves like a
multiperiodic function, i.e. it is a superposition of a large number
of periodic functions with different frequencies. Therefore this
function will retake values arbitrarily close to the initial value for
sufficiently large times. This implies that the apparent loss of
information about the non-diagonal terms reappears if one waits a long
enough time. This problem is usually called ``recurrence of
coherence''.  The characteristic time for these phenomena is
proportional to the factorial of the number of involved
frequencies. Although this time is usually large, perhaps exceeding
the age of the universe, at least in principle it implies that the
measurement process does not correspond to a change from a pure to a
mixed state in a fundamental way.

The above derivation was done using ordinary quantum mechanics in
which one assumes an ideal clock is used to measure time. If one
redoes the derivation using the effective equation we derived for
quantum mechanics with real clocks one gets the same expression for
$z(t)$ except that it is multiplied by $\prod_k\exp\left(-(2 g_k)^2
T^{4/3}_{\rm Planck} t^{2/3}\right)$.  That means that asymptotically
the off diagonal terms indeed vanish, the function $z(t)$ is not
periodic anymore. Although the exponential term decreases slowly with
time, the fact that there is a product of them makes the effect quite
relevant. Therefore we see that including real clocks in quantum
mechanics offers a mechanism to turn pure states into mixed states in
a way that is desirable to explain the problem of measurement in
quantum mechanics.

\section{Directions for future work}

As we showed in the previous section, at least for simple models, the
loss of coherence that arises in quantum mechanics when one considers real
clocks may open a possibility to endow quantum mechanics with a fundamental
process of measurement without the usual conceptual problems. It is clear,
however, that more work is needed before one considers this a definitive
answer to the problem.

The first aspect of this approach that should be further studied is to
consider more realistic models. Examples of such models can be found,
for instance, in the book by Omn\'es \cite{omnes}.

It is clear that to consider only the use of real clocks without
introducing real measuring rods is insufficient to completely address
the issues of the measurement process. d'Espagnat \cite{despa} has
proposed certain observables that are preserved by the unitary
evolution that involve the system and the environment. Such
observables would change drastically in value if the reduced density
matrix were to turn into an proper mixture. Therefore a
measurement on the system plus environment of the observable would
allow to test if the reduced density matrix was in a pure or mixed
state and would be incompatible with any process of reduction of the
wavefunction. Such observables are very difficult to measure since
they imply the simultaneous measurement of a very large number of
microsystems, so for practical purposes again this is not a
problem. It is however, a conceptual objection. In quantum mechanics
with real clocks, the d'Espagnat observables are preserved upon
evolution, since all quantities that commute with the Hamiltonian are
automatically preserved in the evolution with respect to the real
clock. It is to be noted that the measurement of the observable
requires to measure many systems that are in different positions in
space. A proper treatment therefore requires to consider not only the
use of real clocks in quantum mechanics but also the use of ``real
measuring rods''. This is more involved, since it implies dealing with
quantum field theory, and to do it in a way that is not conventional
(in quantum field theory one computes correlation functions and not
the evolution of density matrices). We have carried out a sketch of
the problem in reference \cite{spatial}.  We intend to continue this
analysis in more depth and to relate it to the problem of measurement
and in particular to an analysis of the observables of d'Espagnat.

We also plan to study the impact of the fundamental loss of coherence
in the various interpretations of quantum mechanics. For instance,
in Everitt's relative state (``many worlds'') interpretation, a key
element is that the many worlds all evolve unitarily. If the evolution
is fundamentally non-unitary this interpretation loses its compelling
nature.

It should be noted that even after the inclusion of both temporal and
spatial loss of coherence as argued above, problems still persist with
the interpretation of quantum mechanics. As is well known in
traditional decoherence solutions to the problem, the fact that the
reduced density matrix at the end is in a diagonal form is not
necessarily a completely satisfactory solution to the measurement
problem. This is known as the ``and-or'' problem. As Bell \cite{Bell}
put it ``if one were not actually on the lookout for probabilities,
... the obvious interpretation of even $\rho'$ [the reduced density
matrix] would be that the system is in a state in which various
$|\Psi_m>$'s coexist:
\begin{equation}
|\Psi_1><\Psi_1|\quad {\rm and}\quad |\Psi_2><\Psi_2| 
\quad {\rm and} \quad \ldots
\end{equation}
This is not at all a {\it probability} interpretation, in which the
different terms are seen not as {\it coexisting} but as {\it
alternatives}.''  It is not obvious how our contribution to the
problem changes anything in the discussion of this point.

Even though the proposed research listed above is self-contained and
within the context of ordinary quantum mechanics and quantum field
theory, it is worthwhile mentioning its connection with a broader line
of work taking place in the context of quantum gravity. The authors
have been involved in the last four years in constructing an approach
to quantum gravity called ``consistent discretizations'' (see
\cite{ashtekar} for a recent review). The philosophy of this approach
is similar to that of lattice researchers who deal with QCD. It
consists in discretizing general relativity and using the resulting
discrete theory to probe issues of interest to quantum gravity. The
discrete theories are constructed in such a way that is very
advantageous, in particular the theories are constraint free.  This
allows to work out many proposals that have run into technical
problems in quantum gravity. One such proposal is the use of real
clocks and measuring rods to analyze the theory and therefore ``solve
the problem of time''. The use of relational ideas has long been
sought as a means of dealing properties with the physics of
diffeomorphism invariant theories like general relativity. Up to now
it has always been tampered by technical problems associated with the
presence of the constraints in the canonical theory. The discrete
approach proposed, being constraint free allows to do away with these
technical problems and to formulate the relational description. The
fact that this relational description is central to dealing with
quantum gravity and inspires a point of view that can have
implications for the conceptual problems of ordinary quantum mechanics
brings together two of the most fascinating areas modern theoretical
physics in a remarkably unexpected way.

This work was supported in part by grants NSF-PHY-0244335,
NSF-PHY-0554793, FQXi, CCT-LSU, Pedeciba, and the Horace Hearne Jr.
Institute for Theoretical Physics.

\end{document}